\documentclass[aps,prl,twocolumn,superscriptaddress,longbibliography]{revtex4-2}
\usepackage{graphicx}
\usepackage{amssymb,amsmath,mathtools}
\usepackage{bm}
\usepackage{romannum}
\usepackage{dcolumn}
\usepackage{float}
\usepackage[OT1]{fontenc} 
\usepackage{ulem}
\usepackage{url}
\usepackage{mathrsfs}
\usepackage{slashed,comment}
\usepackage{color}
\usepackage{verbatim}
\usepackage{enumitem}
\usepackage{soul,physics}
\usepackage[driverfallback=dvipdfm]{hyperref}
\hypersetup{pdfpagemode=FullScreen,colorlinks=true,breaklinks,urlcolor=blue,linkcolor=blue,citecolor=blue}

\usepackage{subfigure}
\usepackage{amssymb}
\usepackage{bm}
\usepackage{graphicx}
\usepackage{amsmath,amssymb}
\usepackage{tikz,fp}
\usepackage{tikz-cd}
\usetikzlibrary{arrows}
\usetikzlibrary{intersections}
\usetikzlibrary{shapes.geometric}
\usetikzlibrary{decorations.pathmorphing, patterns,shapes,fixedpointarithmetic}
\usetikzlibrary{decorations.markings}

\begin{document}
  
  \title{Exact Relation Between  Wehrl-R\'enyi Entropy and Many-Body Entanglement }

  \author{Pengfei Zhang}
  \thanks{PengfeiZhang.physics@gmail.com}
  \affiliation{Department of Physics, Fudan University \& State Key Laboratory of Surface Physics, Shanghai, 200438, China}
  \affiliation{Hefei National Laboratory, Hefei 230088, China}

  \author{$\hspace{-4pt}^{\color{blue},\hspace{-1.5pt}~\dagger}$\ Chen Xu}
  \thanks{They contribute equally to this work.}
  \affiliation{School of Physics, Renmin University of China, Beijing, 100872, China}

 \author{Peng Zhang}
  \affiliation{School of Physics, Renmin University of China, Beijing, 100872, China}
  \affiliation{Key Laboratory of Quantum State Construction and Manipulation (Ministry of Education), Renmin University of China, Beijing, 100872, China}

  \date{\today}

  \begin{abstract}
  Quantum entanglement is key to understanding correlations and emergent phenomena in quantum many-body systems. For $N$ qubits (distinguishable spin-$1/2$ particles) in a pure quantum state, many-body entanglement can be characterized by the purity of the reduced density matrix of a subsystem, defined as the trace of the square of this reduced density matrix. Nevertheless, this approach depends on the choice of subsystem. In this letter, we establish an exact relation between the Wehrl-R\'enyi entropy (WRE) $S_W^{(2)}$, which is the 2nd R\'enyi entropy of the Husimi function of the entire system, and the purities of {\it all} possible subsystems. Specifically, we prove the relation 
  $e^{-S_W^{(2)}} = (6\pi)^{-N} \sum_A \Tr({{\hat \rho}_A}^2)$, 
  where $A$ denotes a subsystem with reduced density matrix ${\hat \rho}_A$, and the summation runs over all $2^N$ possible subsystems. Furthermore, we show that the   WRE can be experimentally measured via a  concrete scheme. 
  Therefore, the WRE is a {\it subsystem-independent} and {\it experimentally measurable}  characterization of the overall entanglement in pure states of $N$ qubits. It can be applied to the study of strongly correlated spin systems, particularly those with all-to-all couplings that do not have a natural subsystem division, such as systems realized with natural atoms in optical tweezer arrays or superconducting quantum circuits.  
  We also analytically derive the WRE for several representative many-body states, including Haar-random states, the Greenberger-Horne-Zeilinger (GHZ) state, and the W state.
  \end{abstract}
  
  \maketitle

  {\it \color{blue}Introduction.--} 
  Exploring the rich landscape of quantum phases in many-body systems remains a central focus of modern condensed matter physics. Recent advances have revealed profound insights drawn from quantum information science, shedding light on how complex correlations are established through quantum entanglement. This perspective has significantly deepened our understanding of a wide range of quantum phenomena-from topologically ordered states characterized by long-range entanglement \cite{Kitaev2006topological,Xiaogang2006} to emergent spacetime geometry arising from entanglement structures in holographic systems \cite{Sachdevbook2018,Qixiaoliang2018,Raamsdonk2010,Swingle2012}. Entanglement dynamics has also emerged as a diagnostic tool for distinguishing between thermalized and localized states \cite{Moore2012,Abanin2014,Abanin2014arXiv,Fazio2006,Prelovsek2008}. More recently, the study of quantum evolution under repeated measurements has unveiled a new paradigm, where entanglement serves as an order parameter for dynamical phases \cite{Fisher2018,Nahum2019,Smith2019}. These developments highlight the importance of systematically characterizing entanglement in many-body quantum states. For pure states, entanglement is typically quantified by the von Neumann entropy or the purity of the reduced density matrix of a given subsystem. However, such measures inherently depend on the choice of subsystem and are therefore unfavorable for systems with all-to-all couplings, which lack a natural partitioning.
  
    \begin{figure}[t]
    \centering
    \includegraphics[width=0.85\linewidth]{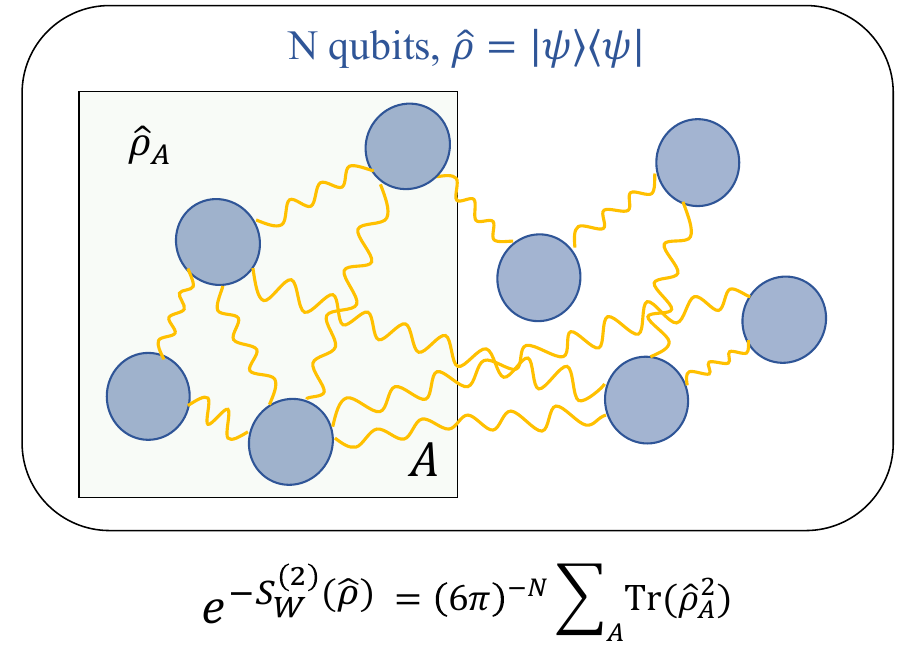}
    \caption{We present a schematic illustrating the relationship between the WRE $S_W^{(2)}$ and the sum of purities over all possible subsystems, thereby confirming that the WRE provides an efficient characterization of the overall entanglement within the entire system.  }
    \label{fig:schemticas}
  \end{figure}

  In this letter, we focus on a system of $N$ qubits (distinguishable spin-$1/2$ prepared in a pure quantum state. For this system, we establish an exact relation between the purities of all subsystems and the Wehrl–R\'enyi entropy (WRE), which is the second R\'enyi entropy of the Husimi function of the entire system (Eq.~(\ref{eqn:relation}), Fig.~\ref{fig:schemticas}). Furthermore, we present an experimental protocol that allows for the direct and efficient measurement of the WRE. These findings demonstrate that the WRE is an effective and experimentally measurable tool for characterizing the overall entanglement of the system.
  Thus, it is particularly useful for studying quantum properties of $N$-quibt systems with all-to-all connectivity, including theoretical models such as Brownian circuits \cite{Lee2008,Peper2013,Lashkari2013,Chen2019}, the Sachdev–Ye–Kitaev model (and its variants) \cite{Kitaev2015,Stanford2016,Suh2018,Sachdev2022}, quantum $p$-spin models \cite{Derrida1981,Gross1984,Crisanti1992,Crisanti1993,Cugliandolo1993,Castellani2005,Bapst2012}, as well as experimental platforms like reconfigurable Rydberg atom arrays \cite{Evered2023,Thompson2023,Bluvstein2024,Lukin2020,Bluvstein2021,Ebadi2022,Lukin2022,Kaufman2023,Endres2024arXiv,Tao2024PRL,Kaufman2024arXiv}.

  {\it \color{blue}Wehrl-R\'enyi Entropy (WRE).--} We prepare the system in a pure state $|\psi\rangle$. Equivalently, the state of the system can be described by the Husimi function \cite{Husimi1940}
  \begin{equation}
  P_H(\hat{\rho},\mathbf{n})=\frac{1}{(2\pi)^N}\bra{\mathbf{n}}\hat{\rho}\ket{\mathbf{n}},
  \end{equation}
  Here, $\hat\rho=|\psi\rangle\langle\psi|$ is the $N$-body density matrix, and $\ket{\mathbf{n}}$ is the SU(2)$^{\otimes N}$ spin coherent state, defined as a direct product of the spin coherent state of each qubit
  \begin{equation}
  \ket{\mathbf{n}}=\ket{\mathbf{n}_1}_1\otimes \ket{\mathbf{n}_2}_2\otimes...\otimes \ket{\mathbf{n}_N}_N,
  \end{equation}
  where $\ket{\mathbf{n}_j}_j$ is the spin coherent state of the $j$-th qubit, with respect to the direction $\mathbf{n}_j\in S^2$. Specifically, we have $[\hat{\bm \sigma}^{(j)}\cdot \mathbf{n}_j]\ket{\mathbf{n}_j}_j=\ket{\mathbf{n}_j}_j$, where $\hat{\bm \sigma}^{(j)}$ is the Pauli operator vector of the qubit $j$. For conciseness, we use $\mathbf{n}$ to denote the collection $\{\mathbf{n}_j\}$. Using the (over) completeness relation for spin coherent states, it is straightforward to show that the Husimi function is properly normalized: $\int d\mathbf{n}~ P_H(\hat{\rho},\mathbf{n})=1$, which is similar as a classical distribution function. A statistical interoperation for the Husimi function was proposed in Ref.~\cite{ChenXu2025}.

 The WRE $S_W^{(2)}$ with respect to state $|\psi\rangle$ is defined as the 2nd R\'enyi entropy of the Husimi function:
   \begin{equation}
  S_W^{(2)}({\hat \rho})\equiv-\ln\left[\int d{{\mathbf n}}\big[P_H({\hat \rho};  {{\mathbf n}})\big]^2\right].\label{re}
   \end{equation}
 As we will see, $S_W^{(2)}({\hat \rho})$ is invariable under any local unitary transformation $\otimes_{j=1}^N\hat u_j$, with $\hat u_j$ being a unitary transformation of qubit $j$. Thus, for pure state $\hat \rho=|\psi\rangle\langle \psi|$, the value of $S_W^{(2)}({\hat \rho})$ is only determined by the entanglement property. Additionally, Eq.~(\ref{re}) implies that the value of $S_W^{(2)}({\hat \rho})$ increases when the Husimi function $P_H({\hat \rho}; {{\mathbf n}})$ becomes more extended in the space of $\mathbf n$, which should originate from quantum entanglement. Therefore, we expect the WRE to characterize the complexity of quantum entanglement in pure many-body states. In previous studies, it has been proposed to use the Wehrl entropy $S_W({\hat \rho})\equiv-\int d{{\mathbf n}}P_H({\hat \rho};  {{\mathbf n}})\ln[P_H({\hat \rho};  {{\mathbf n}})]$, which has the similar behaviors and properties as $S_W^{(2)}({\hat \rho})$, as a measure of entanglement complexity \cite{ChenXu2024}.

{\it \color{blue}Relation between WRE and Subsystem Purities.--} 
  Next, we prove that the WRE, $S_W^{(2)}({\hat \rho})$, satisfies the exact relation
    \begin{equation}\label{eqn:relation}
  e^{-S_W^{(2)}({\hat \rho})}=\frac{1}{(6\pi)^N}\sum_A \Tr({{\hat \rho}_A}^2).
  \end{equation}
  Here, \(A\) denotes a specific subsystem, and \({\hat \rho}_A = \text{Tr}_B[\hat \rho]\) is the reduced density matrix of \(A\), where \(B\) comprises the qubits not included in \(A\). Accordingly, $\Tr({{\hat \rho}_A}^2)$ represents the purity of ${\hat \rho}_A$ \cite{footnote1}. The summation on the right-hand side runs over all $2^N$ possible choices of the subsystem $A$, including the scenario where $A=\varnothing$.  Here we define $\Tr({{\hat \rho}_\varnothing}^2)=1$.

  \begin{figure}[t]
    \centering
    \includegraphics[width=0.95\linewidth]{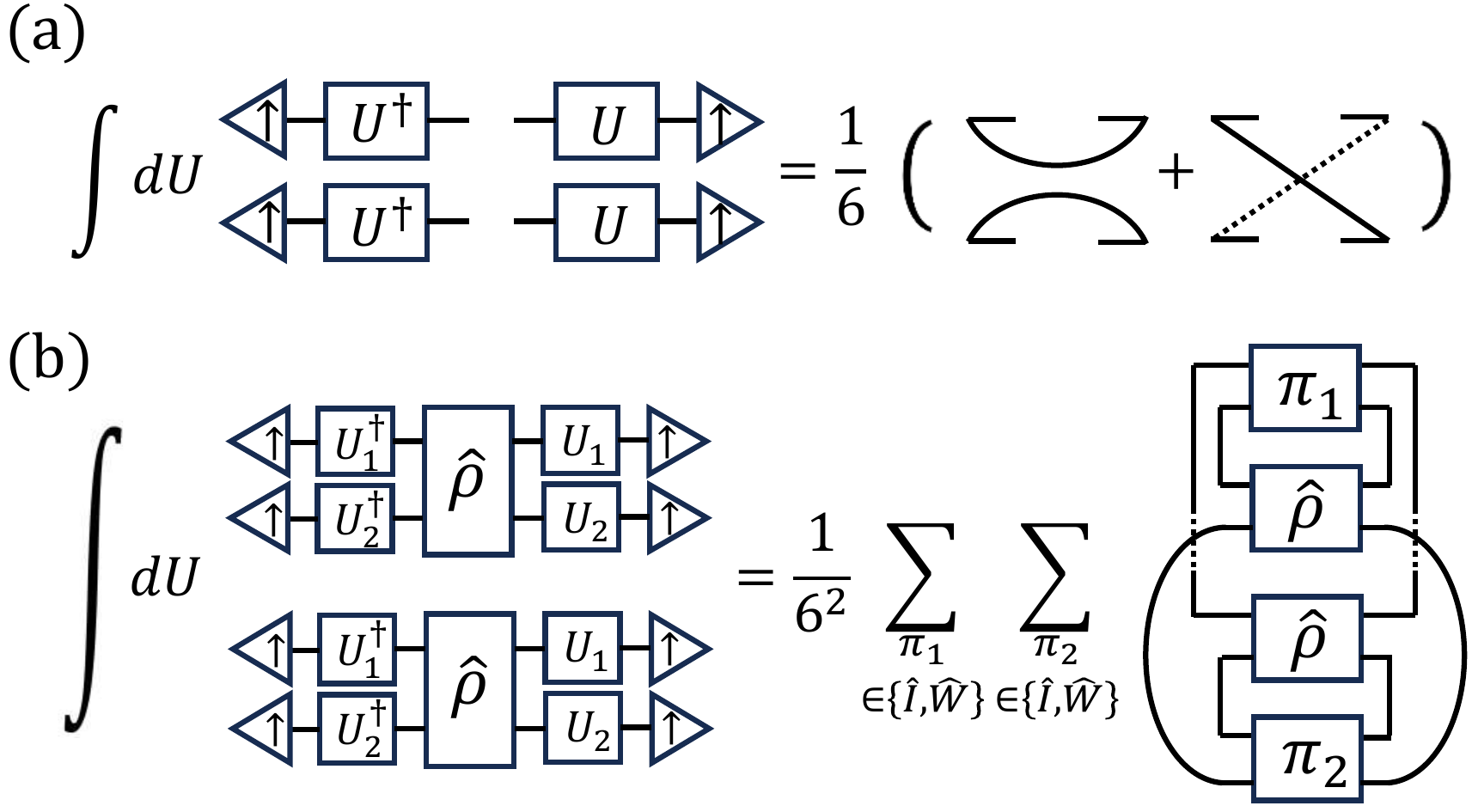}
    \caption{As an illustration, we summarize the key steps in the proof of the exact relation between WRE and subsystem purities. Panel (a) corresponds to Eq.\eqref{eqn:main1}; while panel (b) corresponds to Eq.~\eqref{eqn:main2} with operators $\pi_1,\pi_2\in\{\hat I,\hat W\}$, where $\hat I,\hat W$ represent the identity operator and the SWAP operator respectively.} 
    \label{fig:proof}
  \end{figure}

In Fig.~\ref{fig:proof}, we present a diagrammatic illustration of our approach to proving Eq.~\eqref{eqn:relation}, using the case with $N=2$ as an example.
Below, we outline the main idea and key steps of this approach. Using Eq.(\ref{re}), we first rewrite the left-hand side of Eq.\eqref{eqn:relation} as the integral $\int d{{\mathbf n}}\big[P_H({\hat \rho};  {{\mathbf n}})\big]^2$. We then replace the integral over unit vectors $\mathbf{n}$ with an average over local twirling operations: Any single-qubit spin coherent state $\ket{\mathbf{n}_j}_j$ can be expressed as the SU(2) rotation $\hat{U}_{\mathbf{n}_j}$ acting on $\ket{\uparrow}_j$. 
Here, $\ket{\uparrow}_j$ ($\ket{\downarrow}_j$) denotes the eigenstate of the Pauli operator $\hat\sigma_z^{(j)}$, with eigen-values $+1$($-1$).
It can be shown directly that averaging over $\mathbf{n}_j\in S^2$ is equivalent to averaging over the SU(2) rotations $\hat{U}_{\mathbf{n}_j}$ with respect to the Haar measure, which yields a uniform distribution over the group manifold. This leads to
  \begin{equation}\label{eqn:step1}
  \begin{aligned}
e^{-S_W^{(2)}({\hat \rho})}=\frac{1}{\pi^N}&\int_{\text{T}} d\hat{U} ~ \langle\uparrow\cdots\uparrow|\hat{U}^\dagger{{\hat \rho}}~\hat{U}|\uparrow\cdots\uparrow\rangle ^2.
  \end{aligned}
  \end{equation}
  Here, we have introduced the abbreviation $\hat{U}=\otimes_{j=1}^N \hat{u}_j$ and the subindex $\text{T}$ denotes that each single-qubit unitary $\hat{u}_j$ is sampled over the Haar ensemble independently $\int_{\text{T}} d\hat{U}\equiv \prod_{j=1}^N\int_\text{Haar} d\hat{u}_j$. We have fixed the convention by choosing the normalization that $\int_\text{Haar} d\hat{u}_j=1 $. Note that this expression confirms the invariance of $S_W^{(2)}({\hat \rho})$ under local unitary transformations, as discussed in the previous section.

  Therefore, we need to evaluate the integration over $\hat{U}$. Since rotations on different qubits are independent, the calculation involves the four-point function $\int_{\text{Haar}} d\hat{u}_j~\hat{u}_{j}\otimes\hat{u}_{j}\otimes\hat{u}^\dagger_{j}\otimes\hat{u}^\dagger_{j}$, as in previous studies of quantum many-body chaos \cite{Roberts2017,Kitaev2017arXiv}. For completeness, we provide a detailed explanation in the supplementary material (SM) \cite{SM}. The calculation yields
  \begin{equation} \label{eqn:main1}
   \int_\text{Haar} d\hat{u}_j~(\hat{u}_j\ket{\uparrow})^{\otimes 2} ~^{\otimes 2} {(\bra{\uparrow}\hat{u}_j^\dagger)} = \frac{1}{6}(\hat{I}_j+\hat{W}_j).
  \end{equation}
  Here, $\hat{W}_j$ is the SWAP operation on the $j$-th qubit, defined on the doubled Hilbert space as $\hat{W}_j \ket{s s'}=\ket{s' s}$ for $s,s'\in \{\uparrow,\downarrow\}$. Applying above equation for each single qubit, we obtain 
  \begin{equation}\label{eqn:main2}
  e^{-S_W^{(2)}({\hat \rho})}=\frac{1}{(6\pi)^N}\Tr[\hat\rho\otimes \hat\rho ~\hat{\mathcal{M}}],
  \end{equation}
  with $\hat{\mathcal{M}}=\otimes_{j=1}^N\hat{\mathcal{M}}_j=\otimes_{j=1}^N(\hat{I}_j+\hat{W}_j)$. We further decompose the operator as $\hat{\mathcal{M}}=\sum_A\hat{W}_A$, where each term $\hat{W}_A=\prod_{j\in A}\hat{W}_j$ represents a swap operation acting on a subregion $A$. Finally, Eq.~\eqref{eqn:relation} follows from the identity $\text{Tr}[\hat\rho\otimes \hat\rho~ \hat{W}_A]=\Tr({{\hat \rho}_A}^2)$, which relates the expectation value of the swap operator to the purity of the reduced density matrix on the subsystem $A$.

  At the end of this section, we emphasize that the above analysis remains valid regardless of whether $\hat\rho$ is a pure or mixed state. Consequently, the relation~(\ref{eqn:relation}) holds for the WRE and subsystem purities of both arbitrary pure and arbitrary mixed states. In addition, since the purity satisfies \(2^{-N_A} \leq \Tr({{\hat \rho}_A}^2) \leq 1\), where \(N_A\) denotes the number of qubits in subsystem \(A\), the exact relation~(\ref{eqn:relation}) immediately yields the lower and upper bounds of the WRE, namely,
  \begin{equation}
  N \ln (3\pi) \leq S_W^{(2)}(\hat \rho) \leq N \ln (4\pi).\label{bound}
  \end{equation}

  \begin{figure}[t]
    \centering
    \includegraphics[width=1\linewidth]{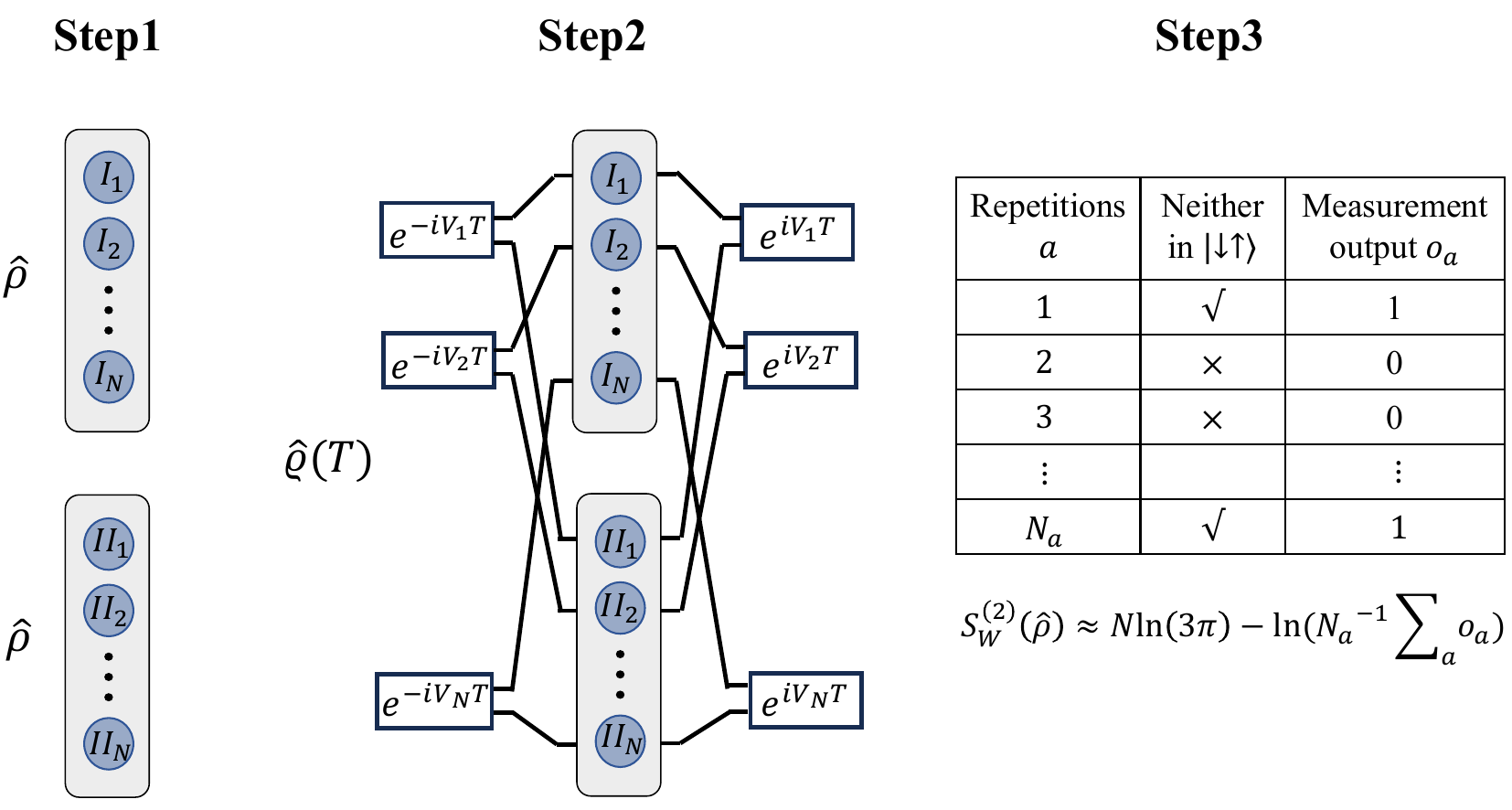}
    \caption{We provide an illustration of the experimental protocol for measuring the WRE. It consists of three steps. {\bf Step 1:} preparing two identical copies of the density matrix. {\bf Step 2:} performing an evolution under the Hamiltonian $\hat{V}$. {\bf Step 3:} performing a projective measurement in the computational basis. More details are presented in the main text. }
    \label{exp}
  \end{figure}

  {\it \color{blue}Experimental Measurement of WRE.--} 
  Inspired by the exact relation in Eq.\eqref{eqn:relation}, we propose an experimental approach to measure the WRE without first determining the Husimi function. The key idea is to reinterpret Eq.\eqref{eqn:main2} as a measurement protocol that can be implemented on two copies of the original physical system. Similar protocols have been proposed for measuring purities \cite{Islam2015,Kaufman2016Science} or R\'enyi-2 correlators \cite{SunNing2025arXiv}. As illustrated in FIG. \ref{exp}, the protocol consists of three steps:

  \textbf{Step 1.} Prepare two identical copies of the many-body system, labeled \Romannum{1} and \Romannum{2}, each in the state $\hat{\rho}$. The full density matrix is then given by $\hat{\varrho}(0)=\hat{\rho}\otimes \hat{\rho}$. 

  \textbf{Step 2.} Turn on an inter-copy coupling $\hat{V}=\sum_{j=1}^{N} \hat{V}_j$ with $\hat{V}_j=\sigma^x_{j,\text{\Romannum{1}}}\sigma^y_{j,\text{\Romannum{2}}}-\sigma^y_{j,\text{\Romannum{1}}}\sigma^x_{j,\text{\Romannum{2}}}$ acting between each pair of qubits on the same site, for a duration $T=\pi/8$.
  Here, $\sigma^{x}_{i,\text{\Romannum{1}/\Romannum{2}}}$ and $\sigma^{y}_{i,\text{\Romannum{1}/\Romannum{2}}}$ are the Pauli operators for qubit $i$ in copy \Romannum{1} or \Romannum{2}.
  As a result of this coupling, the density matrix evolves to $\hat{\varrho}(T)=e^{-i\hat{V}T}\hat\rho\otimes \hat\rho ~e^{i\hat{V}T}$. 

  \textbf{Step 3.} Measure the full system in the $z$ direction (in the computational basis). For each measurement repetition labeled by $a=1,2,...,N_a$, define the outcome $o_a=0$ if any site $j$ is found in the state $\ket{\downarrow}_{j,\text{\Romannum{1}}} \otimes \ket{\uparrow}_{j,\text{\Romannum{2}}}$. Otherwise, we set $o_a=1$. An estimate of the WRE $S_W^{(2)}({\hat \rho})$ is then given by
  \begin{equation}
   S_W^{(2)}({\hat \rho})\approx N\ln3\pi-\ln\Bigg({N_a}^{-1}\sum_{a=1}^{N_a} o_a\Bigg).
  \end{equation}

  As an explanation, Eq.~\eqref{eqn:main2} relates the WRE to the expectation value of the operator $\mathcal{M}$ on the two identical copies of the density matrix $\hat{\varrho}(0)$. Since we have $\mathcal{M} = \otimes_{j=1}^N \mathcal{M}_j$, this requires performing a projective measurement on each pair of qubits at the same site, in the eigenbasis of $\hat{\mathcal{M}}_j = (\hat{I}_j + \hat{W}_j)$. For qubit systems, the singlet state $(\ket{\downarrow}_{j,\text{\Romannum{1}}} \otimes \ket{\uparrow}_{j,\text{\Romannum{2}}}-\ket{\uparrow}_{j,\text{\Romannum{1}}} \otimes \ket{\downarrow}_{j,\text{\Romannum{2}}})/\sqrt{2}$ is the eigenstate of the swap operator $\hat{W}_j$ with eigenvalue $-1$, while the triplet states have eigenvalue $1$. Therefore, if we directly measure all pairs of qubits in the singlet/triplet basis (equivalent to the Bell basis), the measurement outcome of $\mathcal{M}$ is $0$ if any of the outcomes corresponds to a singlet, and $2^N$ otherwise. This additional factor of $2^N$ can be absorbed into the overall normalization in Eq.~\eqref{eqn:main2}. Moreover, the coupling introduced in Step 2 is specifically designed so that the singlet state evolves into the product state $\ket{\downarrow}_{j,\text{\Romannum{1}}} \otimes \ket{\uparrow}_{j,\text{\Romannum{2}}}$. Therefore, we can express Eq. \eqref{eqn:main2} as
  \begin{equation}
  \frac{1}{(6\pi)^N}\Tr[\hat\rho\otimes \hat\rho ~\hat{\mathcal{M}}]=\frac{1}{(3\pi)^N}\Tr[\hat{\varrho}(T)\hat{\mathcal{P}}].
  \end{equation}
 Here, $\hat{\mathcal{P}}$ is a projection operator that yields $1$ if no site is in the state $\ket{\downarrow \uparrow}$ and $0$ otherwise. As a result, the measurement can be performed in the computational basis, which is much more favorable for realistic implementations.

  Finally, we analyze the sample complexity of this measurement protocol. Since the output is either $0$ or $1$, the variance of the measurement outcome is given by
  \begin{equation}
  \begin{aligned}
  \text{Var}(o_a)&=\lim_{N_{a}\gg 1}\left[\frac{1}{N_a}\sum_{a=1}^{N_a}o_a^2-\left(\frac{1}{N_a}\sum_{a=1}^{N_a}o_a\right)^2\right],\\&= e^{-[S_W^{(q)}({\hat \rho})-N\ln 3\pi]}-e^{-2[S_W^{(q)}({\hat \rho})-N\ln 3\pi]}.
  \end{aligned}
  \end{equation}
  Here, we replace the sample average with the expectation value, which is valid for sufficiently large $N_a$. Since the variance of the mean value, ${N_a}^{-1}\sum_{a=1}^{N_a} o_a$, is given by $\text{Var}(o_a)/N_a$, the variance $\text{Var}(o_a)$ effectively characterizes the sample complexity of the protocol. This result is interesting in its own right, as it reveals a direct connection between the entanglement complexity of a quantum state and the number of measurements required to characterize it.

  {\it \color{blue}WRE of Typical States.--} We now present the WRE results for several representative pure states of $N$ qubits. For conciseness, only the final results are shown here, while detailed derivations are provided in the Supplementary Material (SM) \cite{SM}.

 \noindent {\bf Haar random states.} These are random states uniformly sampled from the full Hilbert space, defined as
  \begin{equation}
    |{\rm H}_U\rangle\equiv \hat{U} \ket{\uparrow\uparrow\cdots\uparrow},
  \end{equation}
  where $\hat U$ is a $2^N\times 2^N$ Haar random unitrary matrix in the full Hilbert space. We define the averaged WRE of the Haar random states as
  \begin{equation}
  {S_W^{(2)}(\text{H})}\equiv-\ln\left[\int_{\text{Haar}} d\hat{U}\int d{{\mathbf n}}\big[P_H(|{\rm H}_U\rangle\langle {\rm H}_U|;  {{\mathbf n}})\big]^2\right].
  \end{equation}
  It reflects the WRE of a typical Haar-random state. The closed-form expression can be obtained as
  \begin{equation}
  {S_W^{(2)}(\text{H})}=N\ln(4\pi)-{\ln2}+\ln(1+{2^{-N}}).
  \end{equation}

\noindent   {\bf GHZ state.} The GHZ state  is defined as
  \begin{equation}
    |{\rm GHZ}\rangle\equiv\big[\ket{\uparrow\uparrow\cdots\uparrow}+\ket{\downarrow\downarrow\cdots\downarrow}     \big]/\sqrt{2}.\label{ghz}
  \end{equation}
  The WRE of a this state is:
  \begin{equation}
  {S_W^{(2)}(\text{GHZ})}=N\ln(3\pi)+{\ln2}-{\ln(1+{2^{-N+1}})}.
  \end{equation}
  
\noindent   {\bf W state.} The W state is defined as
  \begin{eqnarray}
    |{\rm W}\rangle=\bigg(\sum_j\hat{\sigma}^-_j\bigg)\ket{\uparrow\uparrow\cdots\uparrow}/\sqrt{N},
  \end{eqnarray}
where $\hat{\sigma}^-_j=|\downarrow\rangle_j\langle\uparrow|$ is the lowering operator of the qubit $j$. The WRE of this state is:
  \begin{equation}
  {S_W^{(2)}({\rm W})}={N}\ln(3\pi)+{\ln2}-{\ln(1+N^{-1})}.
  \end{equation}  

\noindent {\bf $\bm{p}$-Bell state.} The $p$-Bell state contains pair-wised entanglement, which is defined for even $N$ as 
  \begin{equation}
    |p{\text -}{\rm Bell}\rangle\equiv
        \otimes_{j}^{N/2}\big(\ket{\uparrow\uparrow}_{2j-1,2j}+
    \ket{\downarrow\downarrow}_{2j-1,2j}
    \big)/\sqrt{2}.
    \label{h}
  \end{equation}
The WRE of this state is:
  \begin{equation}
     {S_W^{(2)}(p\text{-Bell})}=N\ln(2\sqrt{3}\pi).
  \end{equation}

\begin{figure}[t]
    \centering
    \includegraphics[width=0.95\linewidth]{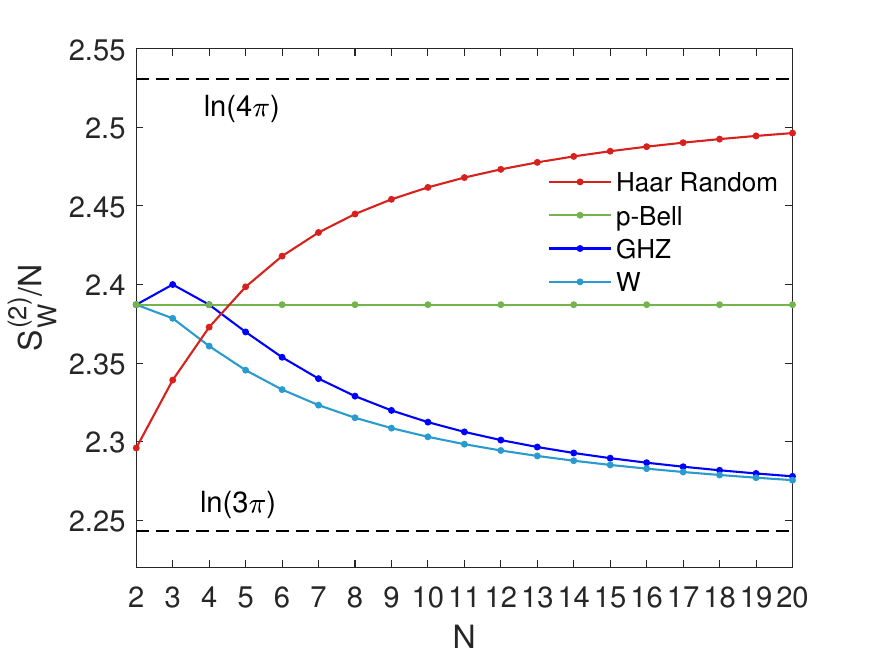}
    \caption{We plot the WRE as a function of the system size $N$ using analytical expressions for representative pure states in many-body systems, including Haar-random states (averaged), the GHZ state, the W state, and the $p$-Bell state. }
    \label{fig:num}
  \end{figure}

 In FIG.~\ref{fig:num}, we illustrate the WRE per particle (i.e., \(S_W^{(2)}/N\)) for the above states as functions of \(N\). It is clear that the behavior of \(S_W^{(2)}/N\) in the limit \(N \rightarrow \infty\) can be classified into following three distinct types, which are similar to those of the Wehrl entropy explored previously \cite{Sugita2003,ChenXu2024}:  
\textbf{(i)} Approaching the upper bound \(\ln (4\pi)\) shown in Eq.~(\ref{bound}). The WRE of Haar-random states exhibits this type of behavior.  
\textbf{(ii)} Approaching the lower bound \(\ln (3\pi)\) shown in Eq.~(\ref{bound}). The WRE of GHZ and W states exhibits this type of behavior.  
 \textbf{(iii)} Approaching a value between the two bounds mentioned above. The WRE of \(p\)-Bell states exhibits this type of behavior. 
These behaviors reflect the distinct overall entanglement properties of various quantum states as $N\rightarrow\infty$.

  {\it \color{blue}Discussions.--} 
  We prove the exact relation~\eqref{eqn:relation} between the WRE of $N$ qubits and traditional measures of entanglement, such as subsystem purities, and further propose a direct experimental protocol for measuring the WRE. In addition, we analytically derive the WRE for several representative pure states, with their behavior in the limit $N \rightarrow \infty$ revealing a clear classification of these states.
    
  Our results show that the WRE can serve as a subsystem-independent characterization of the overall entanglement for many-body pure states of qubits. Therefore, it would be intriguing to investigate the relation between the WRE and other observable  properties of strongly-interacting qubits, such as the ones in the celebrated Sachdev-Ye-Kitaev (SYK) model, which has a holographic duality \cite{Maldacena2016}. It is also interesting to investigate the relationship between the WRE and holographic complexity \cite{Susskind1995,Susskind2016,Susskind2014,Susskind2016PRL,Susskind2016PRD}, which is conjectured to correspond to the bulk action, through the SYK model. Additionally, comparisons with circuit complexity \cite{Susskind2016FP,Susskind2014arXiv,Susskind2016} or Krylov complexity \cite{Parker2019} could also be fruitful.  Furthermore, applying the WRE to diagnose measurement-induced criticality is a promising direction.

  \vspace{5pt}
  \textit{Acknowledgments.}
  We thank Langxuan Chen for helpful discussions. This project is supported by the NSFC under grant 12374477 (Pengfei Zhang), the Shanghai Rising-Star Program under grant number 24QA2700300 (Pengfei Zhang), the Innovation Program for Quantum Science and Technology 2024ZD0300101 (Pengfei Zhang), the National Key Research and Development Program of China (Grant No. 2022YFA1405300) (Peng Zhang), and NSAF Grant No. U1930201 (Peng Zhang).


\begin{thebibliography}{99}  
\bibitem{Kitaev2006topological}A. Kitaev and J. Preskill, Topological entanglement entropy, \href{
		https://doi.org/10.1103/PhysRevLett.96.110404}{Phys. Rev. Lett.  {\bf 96}, 110404 (2006).}
	
	
\bibitem{Xiaogang2006}M. Levin and X.-G. Wen, Detecting Topological Order in a Ground State Wave Function, \href{
		https://doi.org/10.1103/PhysRevLett.96.110405}{Phys. Rev. Lett.  {\bf 96}, 110405 (2006).}	
		

\bibitem{Sachdevbook2018}	S. A. Hartnoll, A. Lucas, and S. Sachdev, Holographic
quantum matter (MIT press, 2018).

		
\bibitem{Qixiaoliang2018}X.-L. Qi, Does gravity come from quantum information?, \href{
	https://doi.org/10.1038/s41567-018-0297-3}{Nature Physics  {\bf 14}, 984-987 (2018).}	
	
\bibitem{Raamsdonk2010}M. Van Raamsdonk, Building up space–time with quantum entanglement, \href{
	https://doi.org/10.1142/S0218271810018529}{Gen. Relativ. Gravit.  {\bf 42}, 2323 (2010).}		
	
\bibitem{Swingle2012}B. Swingle, Entanglement renormalization and holography, \href{https://doi.org/10.1103/PhysRevD.86.065007}{Phys. Rev. D  {\bf 86}, 065007 (2012).}		
	
			
\bibitem{Moore2012}J. H. Bardarson, F. Pollmann, and J. E. Moore, Unbounded growth of entanglement in models of many-body
localization, \href{
	https://doi.org/10.1103/PhysRevLett.109.017202}{Phys. Rev. Lett.  {\bf 109}, 017202 (2012).}	

\bibitem{Abanin2014}M. Serbyn, Z. Papi\'c, and D. A. Abanin, Universal slow
growth of entanglement in interacting strongly disordered
systems, \href{
	https://doi.org/10.1103/PhysRevLett.110.260601}{Phys. Rev. Lett.  {\bf 110}, 260601 (2013).}	

\bibitem{Abanin2014arXiv}I. H. Kim, A. Chandran, D. A. Abanin, Local integrals of
motion and the logarithmic lightcone in many-body local
ized systems, \href{	
	https://doi.org/10.48550/arXiv.1412.3073}{arXiv: 1412.3073 (2014).}	
	
\bibitem{Fazio2006}G. De Chiara, S. Montangero, P. Calabrese, and R. Fazio, Entanglement entropy dynamics of Heisenberg chains, \href{https://doi.org/10.1088/1742-5468/2006/03/P03001}{J.
	Stat. Mech.  {\bf 2006}, P03001 (2006).}		
			

\bibitem{Prelovsek2008}M. \v{Z}nidari\v{c}, T. Prosen, and P. Prelov\v{s}ek
, Many-body
localization in the Heisenberg XXZ magnet in a random
field, \href{
	https://doi.org/10.1103/PhysRevB.77.064426}{Phys. Rev. B  {\bf 77}, 064426 (2008).}	

\bibitem{Fisher2018}Y. Li, X. Chen, and M. P. A. Fisher, Quantum Zeno effect and the many-body entanglement transition, \href{https://doi.org/10.1103/PhysRevB.98.205136}{Phys. Rev. B  {\bf 98}, 205136 (2018).}	

\bibitem{Nahum2019}B. Skinner, J. Ruhman, and A. Nahum, Measurement-Induced Phase Transitions in the Dynamics of Entanglement, \href{https://doi.org/10.1103/PhysRevX.9.031009}{Phys. Rev. X  {\bf 9}, 031009 (2019).}	

\bibitem{Smith2019}A. Chan, R. M. Nandkishore, M. Pretko, and G. Smith, Unitary-projective entanglement dynamics, \href{https://doi.org/10.1103/PhysRevB.99.224307}{Phys. Rev. B  {\bf 99}, 224307 (2019).}	
\bibitem{Lee2008}
J. Lee and F. Peper, On brownian cellular automata, in
Automata (2008) pp. 278–291.

\bibitem{Peper2013}F. Peper, J. Lee, J. Carmona, J. Cortadella, and K. Morita, Towards the fast scrambling conjecture, \href{https://doi.org/10.1145/2422094.2422097}{ACM. J. Emerging
	Technol. Comput. Syst. {\bf 9}, 1 (2013).}	

\bibitem{Lashkari2013}N. Lashkari, D. Stanford, M. Hastings, T. Osborne, and P. Hayden, Towards the fast scrambling conjecture, \href{https://doi.org/10.1007/JHEP04(2013)022}{JHEP {\bf 2013}, 22 (2013).}	

\bibitem{Chen2019}T. Zhou and X. Chen, Operator dynamics in a Brownian quantum circuit, \href{https://doi.org/10.1103/PhysRevE.99.052212}{Phys. Rev. E {\bf 99}, 052212 (2019).}	

\bibitem{Kitaev2015}	Alexei Kitaev, “A simple model of quantum holography,” (2015).

\bibitem{Stanford2016}J. Maldacena and D. Stanford, Remarks on the Sachdev-Ye-Kitaev model, \href{https://doi.org/10.1103/PhysRevD.94.106002}{Phys. Rev. D {\bf 94}, 106002 (2016).}	

\bibitem{Suh2018}A. Kitaev and S. J. Suh, The soft mode in the Sachdev-Ye-Kitaev model and its gravity dual, \href{https://doi.org/10.1007/JHEP05%282018%29183}{JHEP {\bf 05}, 183 (2018).}	

\bibitem{Sachdev2022}D. Chowdhury, A. Georges, O. Parcollet, and
S. Sachdev, Sachdev-ye-kitaev models and beyond: Window into non-fermi liquids, \href{https://doi.org/10.1103/RevModPhys.94.035004}{Rev. Mod. Phys. {\bf 94}, 035004 (2022).}	
\bibitem{Derrida1981}B. Derrida, Random-energy model: An exactly solvable model of disordered systems, \href{https://doi.org/10.1103/PhysRevB.24.2613}{Phys. Rev. B {\bf 24}, 2613 (1981).}	

\bibitem{Gross1984}D. Gross and M. Mezard, The simplest spin glass, \href{https://doi.org/10.1016/0550-3213(84)90237-2}{Nucl. Phys. B {\bf 240}, 431 (1984).}	

\bibitem{Crisanti1992}A. Crisanti and H.-J. Sommers, The Spherical p-Spin
Interaction SpinGlassModel: The Statics, \href{https://doi.org/10.1007/BF01309287}{Z. Phys. B {\bf 87}, 341 (1992).}	


\bibitem{Crisanti1993}A. Crisanti, H. Horner,and H.-J. Sommers, The spherical p-spin interaction spin-glass model, \href{https://doi.org/10.1007/BF01312184}{Z. Phys. B {\bf 92}, 257 (1993).}	

\bibitem{Cugliandolo1993}L. F. Cugliandolo and J. Kurchan, Analytical Solution of the
Off-Equilibrium Dynamics of a Long Range Spin-Glass
Model, \href{https://doi.org/10.1103/PhysRevLett.71.173}{Phys. Rev. Lett. {\bf 71}, 173 (1993).}	

\bibitem{Castellani2005}T. Castellani and A. Cavagna, Spin-Glass Theory for
Pedestrians, \href{https://doi.org/10.1088/1742-5468/2005/05/P05012}{J. Stat. Mech. {\bf 2005}, P05012 (2005).}	

\bibitem{Bapst2012}V. Bapst and G. Semerjian, On quantum mean-field models and their quantum annealing, \href{https://doi.org/10.1088/1742-5468/2012/06/P06007}{J. Stat. Mech.: Theory Exp. {\bf 2012}, P06007 (2012).}	



\bibitem{Evered2023}S. J. Evered, D. Bluvstein, M. Kalinowski, S. Ebadi,
	T. Manovitz, H. Zhou, S. H. Li, A. A. Geim, T. T. Wang,
	et al.,  High-fidelity parallel entangling gates on a neutral-atom quantum computer, \href{https://doi.org/10.1038/s41586-023-06481-y}{Nature {\bf 622}, 268 (2023).}	

\bibitem{Thompson2023}S. Ma, G. Liu, P. Peng, B. Zhang, S. Jandura, J. Claes,
A. P. Burgers, G. Pupillo, S. Puri, and J. D. Thompson,  High-fidelity gates and mid-circuit erasure conversion in
an atomic qubit, \href{https://doi.org/10.1038/s41586-023-06438-1}{Nature {\bf 622}, 279 (2023).}	

\bibitem{Bluvstein2024}D. Bluvstein, S. J. Evered, A. A. Geim, S. H. Li,
H. Zhou, T. Manovitz, S. Ebadi, M. Cain, M. Kalinowski,
D. Hangleiter, et al.,  Logical quantum processor based on
reconfigurable atom arrays, \href{https://doi.org/10.1038/s41586-023-06927-3}{Nature {\bf 626}, 58 (2024).}	

\bibitem{Lukin2020}R. Bekenstein, I. Pikovski, H. Pichler, E. Shahmoon,
S. F. Yelin, and M. D. Lukin, Quantum metasurfaces with atom arrays, \href{https://doi.org/10.1038/s41567-020-0845-5}{Nature Physics {\bf 16}, 676 (2020).}	

\bibitem{Bluvstein2021}D. Bluvstein, A. Omran, H. Levine, A. Keesling, G. Se
meghini, S. Ebadi, T. T. Wang, A. A. Michailidis,
N. Maskara, W. W. Ho, et al., Controlling quantum
many-body dynamics in driven rydberg atom arrays, \href{https://doi.org/10.1126/science.abg2530}{Science {\bf 371}, 1355 (2021).}	

\bibitem{Ebadi2022}S. Ebadi, A. Keesling, M. Cain, T. T. Wang, H. Levine,
D. Bluvstein, G. Semeghini, A. Omran, J.-G. Liu,
R. Samajdar, et al., Quantum optimization of maximum
independent set using rydberg atom arrays, \href{https://doi.org/10.1126/science.abo6587}{Science {\bf 376}, 1209 (2022).}	

\bibitem{Lukin2022}D. Bluvstein, H. Levine, G. Semeghini, T. T. Wang,
S. Ebadi, M. Kalinowski, A. Keesling, N. Maskara,
H. Pichler, M. Greiner, V. Vuleti\'c, and M. D. Lukin,  A quantum processor based on coherent transport of entangled atom arrays, \href{https://doi.org/10.1038/s41586-022-04592-6}{Nature {\bf 604}, 451 (2022).}	

\bibitem{Kaufman2023}J. W. Lis, A. Senoo, W. F. McGrew, F. R\"onchen,
A. Jenkins, and A. M. Kaufman, Midcircuit operations
using the omg architecture in neutral atom arrays, \href{https://doi.org/10.1103/PhysRevX.13.041035}{Phys. Rev. X  {\bf 13}, 041035 (2023).}	

\bibitem{Endres2024arXiv}H. J. Manetsch, G. Nomura, E. Bataille, K. H. Leung,
X. Lv, and M. Endres, A tweezer array with 6100 highly
coherent atomic qubits, \href{	
	https://doi.org/10.48550/arXiv.2403.12021}{arXiv: 2403.12021 (2024).}	

\bibitem{Tao2024PRL}R. Tao, M. Ammenwerth, F. Gyger, I. Bloch, and J. Zeiher, High-fidelity detection of large-scale atom arrays in
an optical lattice, \href{
	https://doi.org/10.1103/PhysRevLett.133.013401}{Phys. Rev. Lett.  {\bf 113}, 013401 (2024).}	

\bibitem{Kaufman2024arXiv}A. Cao, W. J. Eckner, T. L. Yelin, A. W. Young, S. Jan
dura, L. Yan, K. Kim, G. Pupillo, J. Ye, N. D. Oppong,
and A. M. Kaufman, Multi-qubit gates and schr\"odinger
cat states in an optical clock, \href{	
	https://doi.org/10.48550/arXiv.2402.16289}{arXiv: 2402.16289 (2024).}	
\bibitem{Husimi1940}K. Husimi, Some formal properties of the density matrix, \href{
	https://doi.org/10.11429/ppmsj1919.22.4_264}{Proc. Phys. Math. Soc. Japan {\bf 22}, 264 (1940).}

\bibitem{ChenXu2025}C. Xu, Y. Yu, and P. Zhang,  Bayesian interpretation of Husimi function and
Wehrl entropy, \href{
	https://doi.org/10.1088/1367-2630/ada1f1}{Commun. Theor. Phys. {\bf 77}, 095102 (2025).}
	
\bibitem{ChenXu2024}C. Xu, Y. Yu, and P. Zhang,  Wehrl entropy and entangle
ment complexity of quantum spin systems, \href{
	https://doi.org/10.1088/1572-9494/adb947}{New J. Phys. {\bf 26}, 123034 (2024).}
	
\bibitem{footnote1}	Note that the purity is related to the second R\'enyi entropy $S_A^{(2)}$ of $\hat\rho_A$, which is widely used in many studies, through the relation $\Tr({{\hat \rho}_A}^2)=e^{-S_A^{(2)}}$.

\bibitem{Roberts2017}D. A. Roberts and B. Yoshida, Chaos and complexity by design, \href{
	https://doi.org/10.1007/JHEP04(2017)121}{JHEP {\bf 2017}, 121 (2017).}
	
\bibitem{Kitaev2017arXiv}B. Yoshida and A. Kitaev,
Efficient decoding for the Hayden-Preskill protocol, \href{	
	https://doi.org/10.48550/arXiv.1710.03363
	}{arXiv: 1710.03363 (2017).}	

\bibitem{SM}See Supplemental Material for (1) The calculation about the four-point function and (2) The WRE for some typical pure states.

\bibitem{Islam2015}R. Islam, R. Ma, P. M. Preiss, M. Eric Tai, A. Lukin,
M. Rispoli, and M. Greiner, Measuring entanglement en
tropy in a quantum many-body system, \href{https://doi.org/10.1038/nature15750}{Nature {\bf 528}, 77 (2015).}	


\bibitem{Kaufman2016Science}A. M. Kaufman, M. E. Tai, A. Lukin, M. Rispoli,
R. Schittko, P. M. Preiss, and M. Greiner, Quantum thermalization through entanglement in an isolated many
body system, \href{https://doi.org/10.1126/science.aaf6725}{Science {\bf 353}, 794 (2016).}	

\bibitem{SunNing2025arXiv}N. Sun, P. Zhang, and L. Feng, Scheme to Detect the Strong-to-weak Symmetry Breaking
via Randomized Measurements, \href{	
	https://doi.org/10.48550/arXiv.2412.18397}{arXiv: 2412.18397 (2025).}	

\bibitem{Sugita2003}A. Sugita,  Moments of generalized husimi distributions
and complexity of many-body quantum states, \href{
	https://doi.org/10.1088/1572-9494/adb947}{Journal of
	Physics A: Mathematical and General {\bf 36}, 9081 (2003).}
	
\bibitem{Maldacena2016}J. Maldacena, D. Stanford, and Z. Yang, Conformal symmetry and its breaking in two-dimensional nearly anti-de Sitter space, \href{https://doi.org/10.1093/ptep/ptw124}{Prog. Theor. Exp. Phys.  {\bf 2016}, 12C104 (2016).}	




\bibitem{Susskind1995}L. Susskind, The world as a hologram, \href{https://doi.org/10.1063/1.531249}{J. Math. Phys.  {\bf 36}, 6377 (1995).}	

\bibitem{Susskind2016}L. Susskind, Computational complexity and black hole horizons, \href{https://doi.org/10.1002/prop.201500092}{Fortsch. Phys.  {\bf 64}, 44 (2016).}	

\bibitem{Susskind2014}D. Stanford and L. Susskind, Complexity and shock wave
geometries, \href{https://doi.org/10.1103/PhysRevD.90.126007}{Phys. Rev. D  {\bf 90}, 126007 (2014).}	

\bibitem{Susskind2016PRL}A. R. Brown, D. A. Roberts, L. Susskind, B. Swingle, and
Y. Zhao, Holographic Complexity Equals Bulk Action?, \href{https://doi.org/10.1103/PhysRevLett.116.191301}{Phys.
	Rev. Lett.  {\bf 116}, 191301 (2016).}	
	
\bibitem{Susskind2016PRD}A. R. Brown, D. A. Roberts, L. Susskind, B. Swingle, and Y.
Zhao, Complexity, action, and black holes, \href{https://doi.org/10.1103/PhysRevD.93.086006}{Phys. Rev. D  {\bf 93}, 086006 (2016).}	
\bibitem{Susskind2016FP}L. Susskind, Entanglement is not enough, \href{https://doi.org/10.1002/prop.201500095}{Fortsch. Phys.  {\bf 64}, 49 (2016).}	

\bibitem{Susskind2014arXiv}L. Susskind and Y. Zhao, Switchbacks and the Bridge to Nowhere, \href{	
	https://doi.org/10.48550/arXiv.1408.2823}{arXiv: 1408.2823 (2014).}	
	
	
	
\bibitem{Parker2019}D. E. Parker, X. Cao, A. Avdoshkin, T. Scaffidi, and E. Altman, A Universal Operator Growth Hypothesis, \href{https://doi.org/10.1103/PhysRevX.9.041017}{Phys. Rev. X  {\bf 9}, 041017 (2019).}	






	

	
	

	

	

	

	

	
	
	
	
\end{thebibliography}

\newpage
		
		\onecolumngrid
		\newpage
		
		\setcounter{equation}{0}
		\setcounter{figure}{0}
		\setcounter{table}{0}
		\setcounter{secnumdepth}{3}
		\makeatletter
		\renewcommand{\theequation}{S\arabic{equation}}
		\renewcommand{\thefigure}{S\arabic{figure}}
		\renewcommand{\thesection}{S\arabic{section}}
		\renewcommand{\bibnumfmt}[1]{[S#1]}
		\renewcommand{\citenumfont}[1]{S#1}
		\providecommand{\tabularnewline}{\\}

		\begin{center}
			{\large \bf Supplemental Material for \textquotedblleft Exact Relation Between  Wehrl-R\'enyi Entropy and Many-Body Entanglement\textquotedblright}
		\end{center}
		
		\begin{center}
			Pengfei Zhang,$^{1,2}$ Chen Xu,$^{3}$ and Peng Zhang$^{3,4}$
		\end{center}

		\begin{minipage}[]{18cm}
			\small{\it
				\centering $^{1}$Department of Physics, Fudan University \& State Key Laboratory of Surface Physics, Shanghai, 200438, China \\
				\centering $^{2}$Hefei National Laboratory, Hefei 230088, China\\
				\centering $^{3}$School of Physics, Renmin University of China, Beijing, 100872, China\\
		\centering $^{4}$Key Laboratory of Quantum State Construction and Manipulation (Ministry of Education), Renmin University of China, Beijing, 100872, China			
				}
		\end{minipage}
		\vspace{8mm}

\section{Calculation of the four-point function and proof of Eq.~(\ref{eqn:main1})}

In this section, we derive the four-point function
$\int_{\text{Haar}} d\hat{u}_j~\hat{u}_{j}\otimes\hat{u}_{j}\otimes\hat{u}^\dagger_{j}\otimes\hat{u}^\dagger_{j}$
which is utilized in the section titled ``{\it Relation between WRE and Subsystem Purities}" of the main text. Additionally, we provide  a proof of Eq.~(\ref{eqn:main1}) from the main text.

\begin{figure}[H]
	\centering
	\subfigure[]{
		\includegraphics[width=0.4\linewidth]{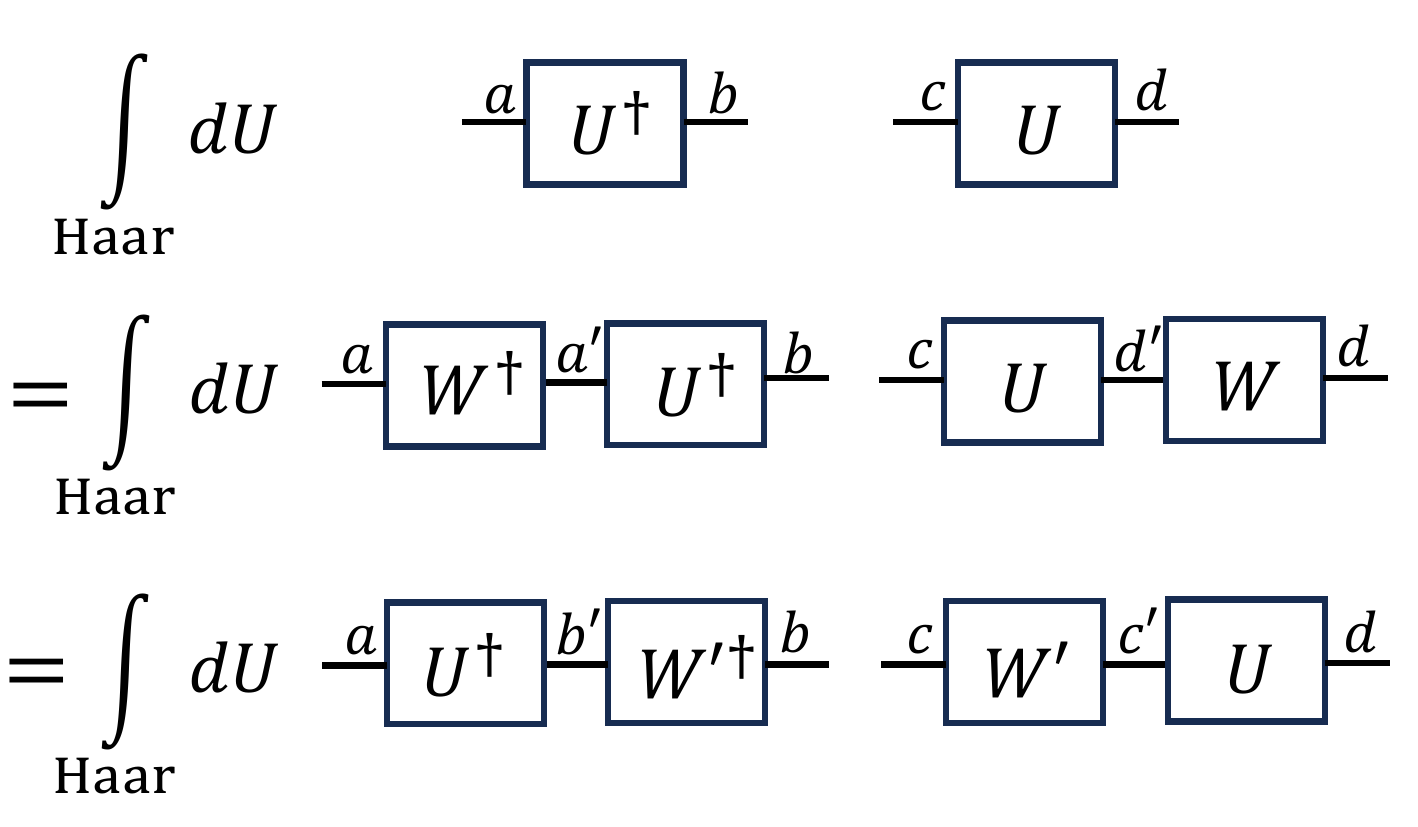}
		\label{Schur1}
	}
	\hspace{1.5cm}
	\subfigure[]{
		\includegraphics[width=0.4\linewidth]{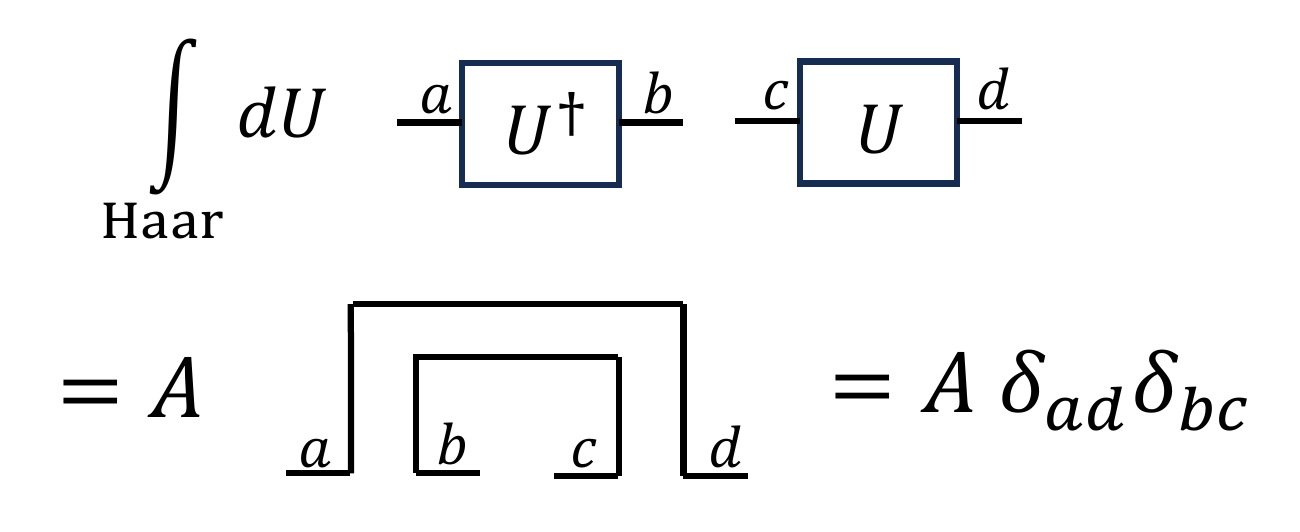}
		\label{Schur2}
	}
	\caption{Calculation of the two-point function. Panel (a) corresponds to Eq.\eqref{WU}; while panel (b) corresponds to Eq.~\eqref{Adelta}.
	}
	\label{Schur}
\end{figure}

To derive the four-point function, we   first calculate the two-point function 
\begin{eqnarray}
I_2(abcd)=\int_{\rm Haar} d\hat{u}_j  (\hat{u}_j^\dagger)_{ab} (\hat{u}_j)_{cd}. \label{tp}
\end{eqnarray}
Notice that the Haar integration $\int_{\rm Haar} d\hat{U}$ for all element $\hat{U}$ in  a unitary group ${\cal U}$ of unitary operators 
satisfies
\begin{eqnarray}\label{relation2}
	\int_{\rm Haar} f(\hat{W}\hat{U})d\hat{U}=\int_{\rm Haar} f(\hat{U}\hat{W})d\hat{U}=\int_{\rm Haar} f(\hat{U})d\hat{U},
\end{eqnarray}
for arbitrary $\hat{W}\in {\cal U}$ and arbitrary function $f$. Due to  this relation, for any $\hat{W}, \hat{W'}\in \text{SU(2)}$, we have (Fig.~\ref{Schur}(a)):
\begin{eqnarray}\label{WU}
	I_2(abcd)=\int_{\rm Haar} d\hat{u}_j  (\hat{W}_j^\dagger\hat{u}_j^\dagger)_{ab} (\hat{u}_j\hat{W}_j)_{cd}=\int_{\rm Haar} d\hat{u}_j  (\hat{u}_j^\dagger\hat{W'}_j^\dagger)_{ab} (\hat{W'}_j\hat{u}_j)_{cd}.
\end{eqnarray}
This result, together with Schur's lemma\rm, yield that (Fig.~\ref{Schur}(b)): 
\begin{eqnarray}\label{Adelta}
	I_2(abcd)= A\delta_{ad}\delta_{bc},
\end{eqnarray}
with $A$ being a to-be-determined coefficient. To find the value of $A$, 
we calculate $\sum_gI_2(aggd)$. For this case, the direct calculation with (\ref{tp}) yields $\sum_gI_2(aggd)=\delta_{ad}$. On the other hand, Eq.~(\ref{Adelta}) directly yields $\sum_g I_2(aggd)=2A\delta_{ad}$. Combining these two results, we finally obtain 
 $A=1/2$, i.e.,
$
	I_2(abcd)= \frac{1}{2}\delta_{ad}\delta_{bc}.
$

\begin{figure}[H]
	\centering
	\subfigure[]{
		\includegraphics[width=0.3\linewidth]{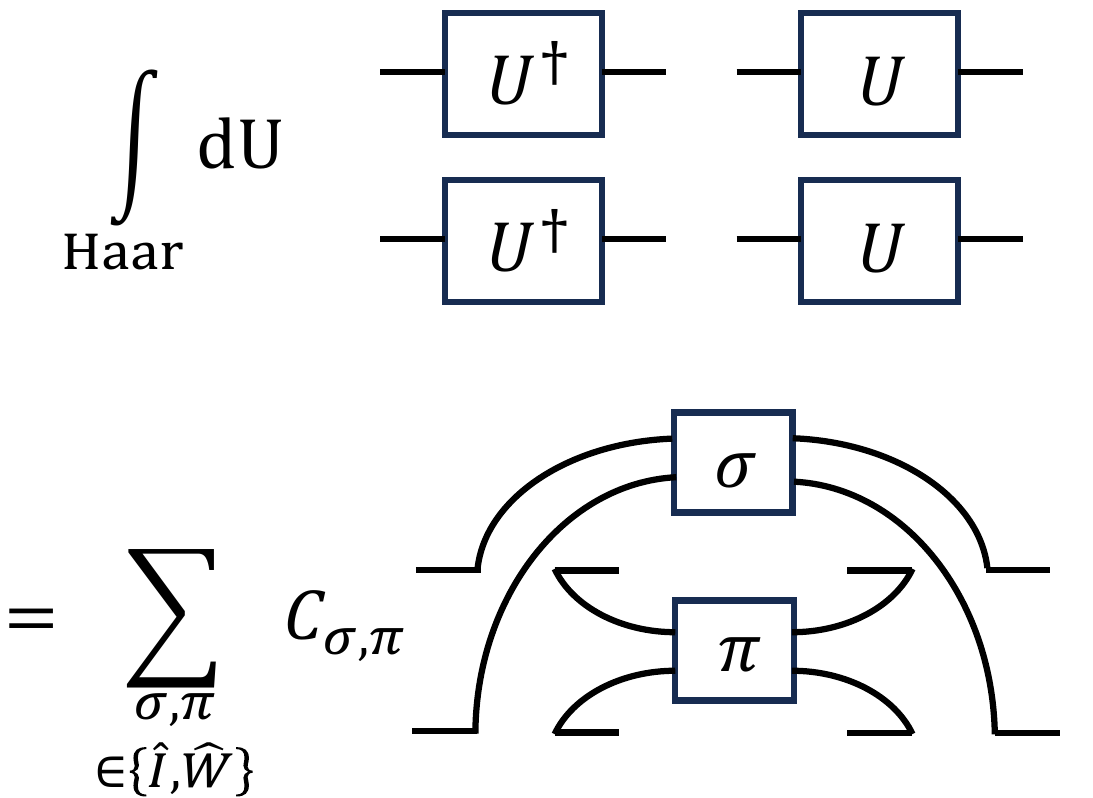}
		\label{Schur_Weylapp}
	}
\hspace{2cm}
	\subfigure[]{
		\includegraphics[width=0.23\linewidth]{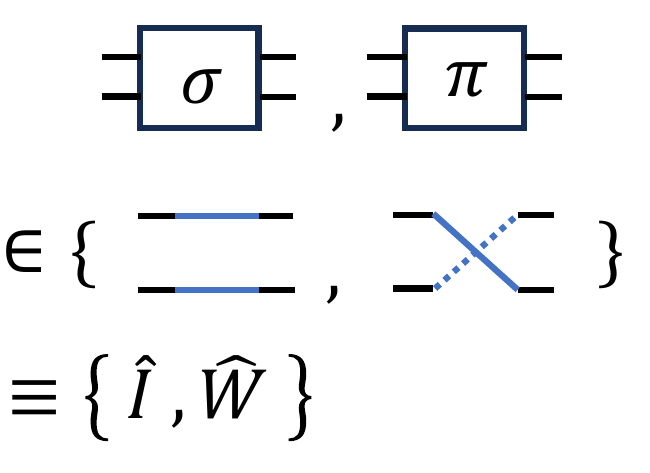}
		\label{sigma_pi}
	}
	\caption{The illustration of the calculation of the general four-point function 
		$\int_{\text{Haar}} d\hat{u}_j~\hat{u}_{j}\otimes\hat{u}_{j}\otimes\hat{u}^\dagger_{j}\otimes\hat{u}^\dagger_{j}$.
		Panel (a) shows the result of the four-point function Haar integration; while panel (b) shows the operators $\{\hat I,\hat W\}$.
	}
	\label{Schur_sigmasupp}
\end{figure}

Next, we calculate the general four-point function $\int_{\text{Haar}} d\hat{u}_j~\hat{u}_{j}\otimes\hat{u}_{j}\otimes\hat{u}^\dagger_{j}\otimes\hat{u}^\dagger_{j}$. 
Using Schur-Weyl duality \cite{Roberts2017}, we can generalize Eq.~(\ref{Adelta}) to the result presented in Fig.~\ref{Schur_sigmasupp}(a). Note that, in contrast to the single coefficient \( A \) in Eq.~(\ref{Adelta}), we now have four coefficients $ C_{\sigma, \pi} $ to be determined, with $\sigma,\pi\in\{\hat I,\hat W\}$. Here $\hat I,\hat W$ represent the identity operator and the SWAP operator, respectively.

\begin{figure}[H]
	\centering
	\includegraphics[scale=0.45]{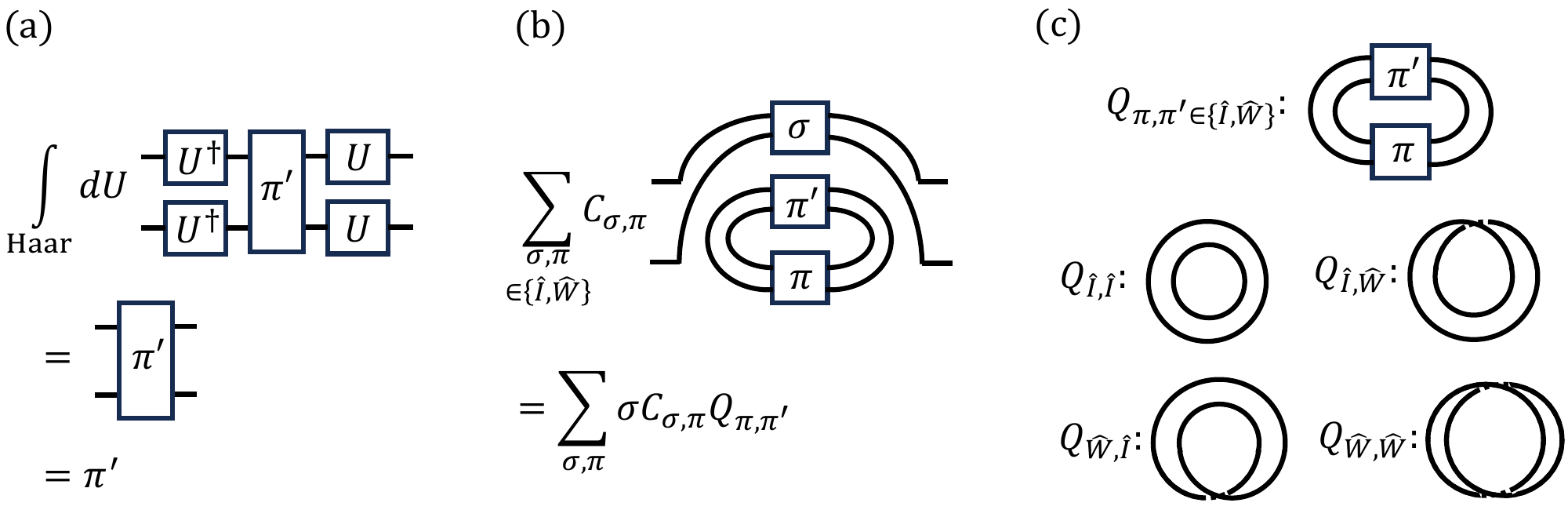}
	\caption{Calculation of the integration $I_4$. Panel (a) corresponds to Eq.\eqref{r1}; while panel (b) corresponds to Eq.~\eqref{r2}.
		 Panel (c) is the calculation for the coefficients $Q_{\pi,\pi'}$ ($\pi,\pi'\in \{\hat I, \hat W\}$) . A closed circle in the figure represents $\Tr(\hat I)\equiv d$.}\label{Weyl03}		
\end{figure}

Furthermore, in order to obtain the coefficient $C_{\sigma,\pi}$, we 
calculate $\mathcal{E}({\pi^\prime})\equiv \int_{\text{Haar}} d\hat{u}_j~\hat{u}_{j}\otimes\hat{u}_{j}~{\pi^\prime}~\hat{u}^\dagger_{j}\otimes\hat{u}^\dagger_{j}$ for $\pi^\prime\in \{\hat I,\hat W\}$.
It is straightforward to verify that $\hat{u}_{j}$ and $\hat{u}_{j}^\dagger $ cancels out. As a result, the set $I_4\equiv\left\{\mathcal{E}(\pi^\prime)\bigg\vert\ \pi^\prime\in\{\hat I, \hat W\}\right\}$ is nothing  
but $\{\hat I, \hat W\}$, i.e.,
\begin{eqnarray}
	I_4\equiv\left\{\mathcal{E}(\pi^\prime)\bigg\vert\ \pi^\prime\in\{\hat I, \hat W\}\right\}=\{\hat I, \hat W\}.\label{r1}
\end{eqnarray}
This is illustrated in Fig.~\ref{Weyl03}(a). Alternatively, we can use the result obtained after performing the Haar integration (the lower term in Fig.~\ref{Schur_sigmasupp}(a)) with the insertion of the operator $\pi'\in\{\hat I, \hat W\}$, and the calculation yields (see in Fig.~\ref{Weyl03}(b)):
\begin{eqnarray}
	I_4=\left\{\sum_{\sigma,\pi\in \{\hat I,\hat W\}}\sigma C_{\sigma,\pi}Q_{\pi,\pi'}
	\Bigg\vert \ \pi^\prime\in\{\hat I,\hat W\}
	\right\}.\label{r2}
\end{eqnarray}
Moreover, as  calculated in
Fig.~\ref{Weyl03}(c), 
 the coefficients $Q_{\pi,\pi'}$ ($\pi,\pi'\in \{\hat I, \hat W\}$) are given by
 \begin{eqnarray}
	\mqty(Q_{\hat I,\hat I}&Q_{\hat I,\hat W} \\Q_{\hat W,\hat I} &Q_{\hat W,\hat W})=\mqty(d^2&d \\d &d^2),
\end{eqnarray}
where $d$ is the
dimension of the Hilbert space of the operator $\hat{ u}_j$, i.e., $d=2$ for our system.
On the other hand,
Eqs~(\ref{r1}, \ref{r2}) yield
$
	\sum_{\pi\in\{\hat I,\hat W\}}C_{\sigma,\pi}Q_{\pi,\pi'}=\delta_{\sigma,\pi'}
$, i.e., $CQ=\hat I$, where
$\hat I$ is the identity operator. Thus the coefficients $C_{\sigma,\pi}$ are
\begin{eqnarray}
	\mqty(C_{\hat I,\hat I}&C_{\hat I,\hat W} \\C_{\hat W,\hat I} &C_{\hat W,\hat W})=Q^{-1}=\mqty(\frac{1}{d^2-1}&\frac{-1}{d(d^2-1)} \\\frac{-1}{d(d^2-1)} &\frac{1}{d^2-1}).\label{cpi}
\end{eqnarray}


\begin{figure}[H]
	\centering
	\includegraphics[scale=0.35]{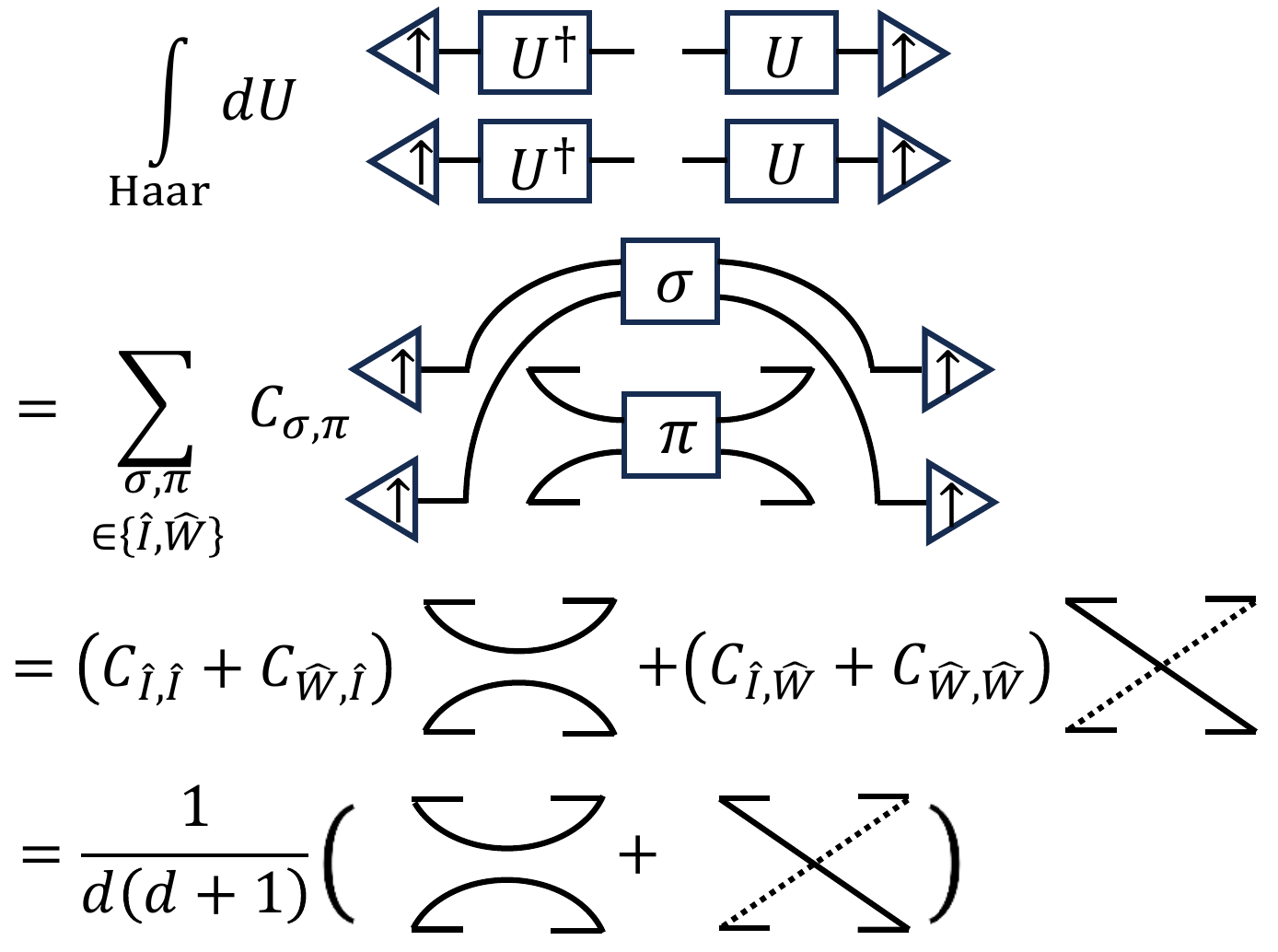}
	\caption{The illustration about the calculation of Eq.~(\ref{s9}).}\label{sketch2}		
\end{figure}

Substituting the coefficients $C_{\sigma,\pi}$  given by Eq.~(\ref{cpi}) into the result of Fig.~\ref{Schur_sigmasupp}(a), we finally obtain
 the specific four-point function (Fig.~\ref{sketch2}):
\begin{eqnarray}\label{resultf}
	\tilde{I}_4(abcd)&=&\int_{\rm Haar} d\hat{u}_j  \left[\langle \uparrow |\hat{u}_j^\dagger\right]_{a}  \,_{b}\left[\hat{u}_j|\uparrow \rangle \right] \,
\left[\langle \uparrow |\hat{u}_j^\dagger\right]_{c} \,
_{d}\left[\hat{u}_j|\uparrow \rangle \right],\nonumber\\
&=& \frac{1}{d(d+1)}(\delta_{ab}\delta_{cd}+\delta_{ad}\delta_{bc}).\label{s9}
\end{eqnarray}
where $d$ is the
dimension of the Hilbert space of the operator $\hat{ u}_j$, as mentioned above.  Thus, the coefficient $\frac{1}{d(d+1)}=\frac1 6$. 
Eq.~(\ref{resultf}) can also be expressed as $
	\int_\text{Haar} d\hat{u}_j~(\hat{u}_j\ket{\uparrow})^{\otimes 2} ~^{\otimes 2} {(\bra{\uparrow}\hat{u}_j^\dagger)} = \frac{1}{6}(\hat{I}_j+\hat{W}_j)$, which is precisely the Eq.~(\ref{eqn:main1}) from the main text.


\section{The WRE of some typical pure states}
In this section, we calculate the WRE of some typical pure states,
which are utilized in the section titled ``{\it WRE of Typical States}" of the main text. For convenience, we define  $M^{(2)}({\hat \rho})$ as the second moment of the Husimi function $P_H({\hat \rho};  {{\mathbf n}})$ of a $N$-body state $\hat \rho$, i.e.,
\begin{eqnarray}
	M^{(2)}({\hat \rho})\equiv\int d{{\mathbf n}}\big[P_H({\hat \rho};  {{\mathbf n}})\big]^2.
\end{eqnarray}
Thus the WRE of the state $\hat \rho$ is related to $M^{(2)}({\hat \rho})$ via
\begin{eqnarray}
S_W^{(2)}({\hat \rho})=-\ln M^{(2)}({\hat \rho}).\label{sw2}
\end{eqnarray}

\noindent{\bf Haar random state.}

\begin{figure}[h]
		\centering
		\includegraphics[width=0.5\linewidth]{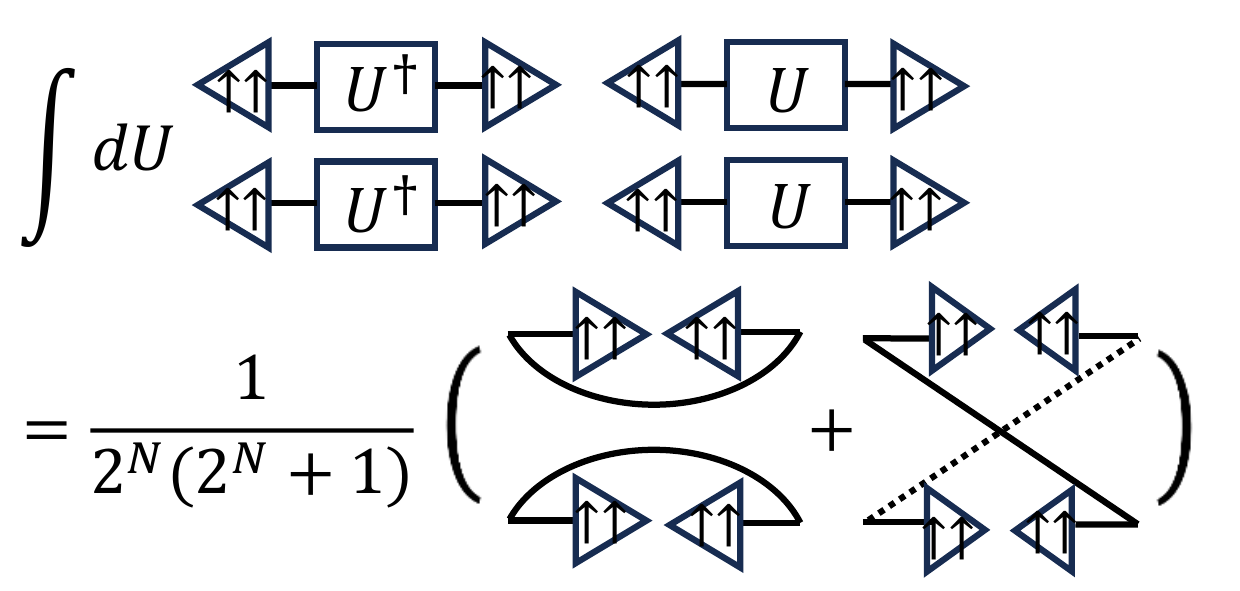}
		\caption{The calculation of $\int_{\text{Haar}} d\hat{U}
		 M^{(2)}\bigg(|{\rm H}_U\rangle\langle {\rm H}_U|\bigg)$. Here the state $|\uparrow\uparrow\rangle$ in this illustration means the  product state of N particles $|\uparrow\uparrow\cdots\uparrow\rangle$ while the operator $U$ means $(\hat{U}^\dagger\otimes_{j=1}^N \hat{u}_j)$. Notice that $d$ is the
		 dimension of the Hilbert space of the operator $U$, i.e., $d=2^N$.}
		\label{fig:Haar}
	\end{figure}

	The Haar random state is defined as
	\begin{equation}
		|{\rm H}_U\rangle\equiv \hat{U} \ket{\uparrow\uparrow\cdots\uparrow},
	\end{equation}
	where $\hat{U}$ is a $2^N\times 2^N$ Haar random unitrary matrix in the full Hilbert space. 
	Here we focus on the averaged WRE of the Haar random states, which is define as
\begin{eqnarray}
	{S_W^{(2)}(\text{H})}\equiv-\ln \left[\int_{\text{Haar}} d\hat{U}
		 M^{(2)}\bigg(|{\rm H}_U\rangle\langle {\rm H}_U|\bigg)\right].
		 \end{eqnarray}
To derive ${S_W^{(2)}(\text{H})}$, we first calculate $\int_{\text{Haar}} d\hat{U}
		 M^{(2)}\bigg(|{\rm H}_U\rangle\langle {\rm H}_U|\bigg)$
	with the approach  similar that for  Eq. \eqref{eqn:main1} (Fig.~\ref{fig:Haar}):
	\begin{eqnarray}
		 & &\int_{\text{Haar}} d\hat{U}
		 M^{(2)}\bigg(|{\rm H}_U\rangle\langle {\rm H}_U|\bigg)
\nonumber\\
		&=&\frac{1}{\pi^N}\int_{\text{Haar}} d\hat{U} \prod_{j=1}^N\int_\text{Haar} d\hat{u}_j\langle\uparrow\uparrow\cdots\uparrow|\otimes_{j=1}^N \hat{u}_j^\dagger\hat{U}|\uparrow\uparrow\cdots\uparrow\rangle^2
		\langle\uparrow\uparrow\cdots\uparrow|\hat{U}^\dagger\otimes_{j=1}^N \hat{u}_j|\uparrow\uparrow\cdots\uparrow\rangle^2,\nonumber\\
		&=&\frac{1}{\pi^N}\int_{\text{Haar}} d\hat{U}\langle\uparrow\uparrow\cdots\uparrow|\hat{U}|\uparrow\uparrow\cdots\uparrow\rangle^2
		\langle\uparrow\uparrow\cdots\uparrow|\hat{U}^\dagger|\uparrow\uparrow\cdots\uparrow\rangle^2,\nonumber\\
		&=&\frac{1}{\pi^N}\frac{2}{ 2^N (2^N+1)}.
	\end{eqnarray}
This result directly yields that the averaged WRE of the Haar random states is
\begin{eqnarray}
	{S_W^{(2)}(\text{H})}=N\ln(4\pi)-{\ln2}+\ln(1+{2^{-N}}).
\end{eqnarray}

	\bigskip

\noindent{\bf GHZ state.}

The GHZ state is define as
	\begin{eqnarray}
		|{\rm GHZ}\rangle\equiv\big[\ket{\uparrow\uparrow\cdots\uparrow}+\ket{\downarrow\downarrow\cdots\downarrow}     \big]/\sqrt{2}.
	\end{eqnarray}
	For arbitrary subsystem $A$ with $0<N_A<N$, the purity is $1/2$. When $N_A=0$ or $ N$, the purity is 1.
	From the exact relation (Eq.~(\ref{eqn:relation})), it follows that
	\begin{eqnarray}
		M^{(2)}(\text{GHZ})=\frac{1}{(6\pi)^N}\left(\sum_{N_A=1}^{N-1} \frac 12 C_N^{N_A}+2\right)=\frac{1}{(6\pi)^N}(1+2^{N-1} ).
	\end{eqnarray}
Substituting this result into Eq.~(\ref{sw2}), we finally obtain:
	\begin{equation}
		{S_W^{(2)}(\text{GHZ})}=N\ln(3\pi)+{\ln2}-{\ln(1+{2^{-N+1}})}.
	\end{equation}
	\bigskip
	
\noindent {\bf W state.}

The W state is defined as
	\begin{eqnarray}
		|{\rm W}\rangle=\bigg(\sum_j\hat{\sigma}^-_j\bigg)\ket{\uparrow\uparrow\cdots\uparrow}/\sqrt{N},
	\end{eqnarray}
	where $\hat{\sigma}^-_j=|\downarrow\rangle_j\langle\uparrow|$ is the lowering operator of the qubit $j$. 
For this state, the reduced density matrix for a generic subsystem $A$ contains two non-zero eigenvalues $N_A/N$ and $1-N_A/N$. From Eq.~(\ref{eqn:relation}), we have
	\begin{eqnarray}
		M^{(2)}(\rm W)=\frac{1}{(6\pi)^N}\sum_{N_A=0}^{N}  C_N^{N_A}\left[\left(\frac{N_A}{N}\right)^2+\left(\frac{N-N_A}{N}\right)^2\right]=\frac{1}{(3\pi)^N}\frac{N+1}{2N},
	\end{eqnarray}
	where $C_N^{N_A}=\frac{N!}{N_A!(N-N_A)!}$.
Substituting this result into Eq.~(\ref{sw2}), we finally obtain:
	\begin{equation}
		{S_W^{(2)}({\rm W})}={N}\ln(3\pi)+{\ln2}-{\ln(1+N^{-1})}.
	\end{equation}
	
	\bigskip
	
\noindent{\bf p-Bell state.} 

The $p$-Bell state is defined as:
	\begin{equation}
		|p{\text -}{\rm Bell}\rangle\equiv
		\otimes_{j}^{N/2}\big(\ket{\uparrow\uparrow}_{2j-1,2j}+
		\ket{\downarrow\downarrow}_{2j-1,2j}
		\big)/\sqrt{2}.
		\label{h}
	\end{equation}
With direct calculation, we find that
	\begin{equation}
	M^{(2)}(p\text{-Bell})=\big[M^{(2)}(\text{GHZ})|_{N=2}\big]^{N/2}=\frac{1}{(2\sqrt{3}\pi)^N}.
	\end{equation}
Substituting this result into Eq.~(\ref{sw2}), we finally obtain:
	\begin{equation}
		{S_W^{(2)}(p\text{-Bell})}=N\ln(2\sqrt{3}\pi).
	\end{equation}




\end{document}